Gregory M. Dickinson*

# An Empirical Study of Obstacle Preemption in the Supreme Court

## TABLE OF CONTENTS

## I. INTRODUCTION

The Supreme Court's preemption jurisprudence over the last few decades has been unpredictable to say the least. Preemption defies traditional conservative–liberal alignment, as conservatives are torn between support of federalism and capitalist efficiency, and liberals are torn between support of strong national governance and multiplicity of legal remedies. As a result, justices often "flip" to take positions in tension with those that they take in federalism cases that do not involve questions of preemption. This tension between competing values combined with the complexity of preemption doctrine and the sheer number of values at stake in preemption cases produces great uncertainty. Court watchers and business interests alike have been unable to predict the path of the Court's jurisprudence.[1]

---

    * Associate, Ropes & Gray, LLP; J.D., Harvard Law School; B.S., Houghton College, Computer Science. Thanks to Matthew Stephenson for helpful comments.

With its recent decision in *Wyeth v. Levine*,[2] however, the Court has clarified its preemption analysis. Justice Thomas, in his concurrence, firmly rejects an entire line of the Court's preemption decisions—"purposes and objectives" obstacle preemption. This rejection is not a sudden development; Justice Thomas has disfavored obstacle preemption for some time. What is significant, though, is the opinion's frankness and the clarity that it lends to the Court's past (and future) conflict preemption analysis.

While urging the Court to abandon its obstacle preemption doctrine as overly broad and unsupported by congressional authorization, Justice Thomas simultaneously advocates for the expansion of what has traditionally been a very narrow category of preemption: impossibility preemption. Such an expansion is likely made necessary by the gap that would otherwise be left by his rejection of the obstacle preemption category. The natural question, then, is whether Justice Thomas's expanded impossibility preemption might look very much like the narrow version of obstacle preemption that has been applied by the Court's liberal bloc for some time. If this is the case, do we now have a firm five-Justice majority applying a unified doctrine to obstacle preemption cases? If so, what does that doctrine look like?

This Article sets out to answer these questions through an empirical analysis of the Court's obstacle preemption decisions. Justice Thomas's drift to disfavor obstacle preemption has been gradual. Although he did not formally renounce the doctrine until *Wyeth*, he has presumably been applying a similar analysis for some time. An examination of each of the Court's obstacle preemption cases over the last fifteen years confirms that presumption. The analysis shows that Justice Thomas's decisions very closely parallel those of the Court's liberal bloc over the same time period. His broad impossibility analysis is thus functionally coterminous with a narrow version of obstacle preemption. Justice Thomas and the Court's liberals form a distinct and somewhat reliable anti-obstacle preemption bloc. The analysis also shows, however, that because Justice Thomas and the Court's liberals arrive at their positions through quite different analyses, their bloc is subject to fracture in certain particularly contentious cases.

Part II presents a brief overview of the Court's current preemption doctrine. Part III details the breakdown of traditional left–right voting blocs in preemption cases and the consequent failure of political voting models to explain the Court's preemption jurisprudence. Part

---

1. Mark D. Rosen, *Contextualizing Preemption*, 102 Nw. U. L. Rev. 781, 781 (2008).
2. 129 S. Ct. 1187 (2009).

IV uses the recent *Wyeth v. Levine* decision as a lens through which to view the Court's peculiar voting patterns in obstacle preemption cases. Part V presents an empirical analysis of the Court's obstacle preemption cases over the last fifteen years. The data show an emerging five-Justice voting bloc opposed to obstacle preemption resulting from the surprising alignment of Justice Thomas with the Court's liberal wing. Finally, Part VI identifies the factors that the loosely aligned voting bloc finds most relevant to its preemption decisions, exposes weak points in the bloc where fracture is likely, and discusses how recent appointments will affect voting alignments.

## II.  DEFINING FEDERAL PREEMPTION

Congress's power of preemption, rooted in the Supremacy Clause of the Constitution,[3] permits federal law to trump state law where it is undesirable or impossible for two independent legal regimes to coexist. The Supreme Court has recognized two primary categories of preemption: express and implied.[4] Express preemption occurs where a federal statute expressly withdraws from the states regulatory power over a certain area of law.[5] Express preemption doctrine, therefore, involves the difficult but familiar judicial task of determining the intended preemptive reach of statutory language.[6] Implied preemption is subdivided into two types: field preemption and conflict preemption.[7] Field preemption occurs where a federal regulatory regime is so pervasive as to imply that Congress intended to occupy an entire field of the law, leaving no room for states to supplement that federal regulation.[8] Similarly, but on a smaller scale, conflict preemption occurs

---

3.  U.S. Const. art. VI, cl. 2 ("This Constitution, and the Laws of the United States . . . shall be the supreme Law of the Land; and the Judges in every State shall be bound thereby, any Thing in the Constitution or Laws of any State to the Contrary notwithstanding.").

4.  *See* Gade v. Nat'l Solid Wastes Mgmt. Ass'n, 505 U.S. 88, 98 (1992) ("Pre-emption may be either expressed or implied, and is compelled whether Congress' command is explicitly stated in the statute's language or implicitly contained in its structure and purpose.") (citations omitted) (internal quotation marks omitted).

5.  *See* Caleb Nelson, *Preemption*, 86 Va. L. Rev. 225, 226–27 (2000).

6.  *See id.*; Daniel E. Troy & Rebecca K. Wood, *Federal Preemption at the Supreme Court*, 2007–2008 Cato Sup. Ct. Rev. 257, 258–60.

7.  *Gade*, 505 U.S. at 98 ("Absent explicit pre-emptive language, we have recognized at least two types of implied pre-emption: field pre-emption, where the scheme of federal regulation is 'so pervasive as to make reasonable the inference that Congress left no room for the States to supplement it,' and conflict pre-emption, where 'compliance with both federal and state regulations is a physical impossibility,' or where state law 'stands as an obstacle to the accomplishment and execution of the full purposes and objectives of Congress.'") (citations omitted).

8.  *See* Cipollone v. Liggett Grp., Inc., 505 U.S. 504, 516 (1992) ("In the absence of an express congressional command, state law is pre-empted if that law actually conflicts with federal law, or if federal law so thoroughly occupies a legislative field 'as to make reasonable the inference that Congress left no room for the States to where, though Congress has demonstrated no intent to occupy an entire field of law, federal law conflicts with a particular state law.[9]  This conflict may take either of two forms.  First, state law will be preempted "where it is impossible for a private party to comply with both state and federal law."[10]  Second, state law will also be preempted where, though it is not literally impossible to comply with both state and federal law, state law "stands as an obstacle to the accomplishment and execution of the full purposes and objectives of Congress."[11]  This taxonomy of preemption yields four fundamental varieties: express preemption, field preemption, impossibility preemption, and obstacle preemption.  This Article will focus primarily on conflict preemption (both impossibility and obstacle preemption), but the others will receive passing attention where relevant to the conflict preemption inquiry.

## III.  JUDICIAL ALIGNMENT ON PREEMPTION

As many studies have shown, political ideology is an important determinant of Supreme Court decisions.  Justices' votes can be explained, at least in part, by their political preferences.[12]  In a typical federalism case, for instance, conservative justices tend to favor states' rights, while more liberal justices tend to favor a strong central government.[13]  In the preemption context, however, political ideology often pulls in opposite directions.[14]  A decision against preemption in favor of states' rights, typically considered conservative, may have a liberal outcome, and vice versa.  "[A] 'liberal' vote for the federal gov-

---

supplement it.'") (quoting Rice v. Santa Fe Elevator Corp., 331 U.S. 218, 230 (1947)); English v. Gen. Elec. Co., 496 U.S. 72, 79 (1990) ("[I]n the absence of explicit statutory language, state law is pre-empted where it regulates conduct in a field that Congress intended the Federal Government to occupy exclusively."); *see also* Nelson, *supra* note 5, at 227 ("The [Supreme] Court has indicated that a federal regulatory scheme may be so pervasive as to imply that Congress left no room for the states to supplant it.") (citations omitted) (internal quotation marks omitted).

9.  *See* Nelson, *supra* note 5, at 227–28.

10.  Crosby v. Nat'l Foreign Trade Council, 530 U.S. 363, 372–73 (2000); *see, e.g.*, Fla. Lime & Avocado Growers, Inc. v. Paul, 373 U.S. 132, 142–43 (1963).

11.  Hillsborough Cnty. v. Automated Med. Labs., Inc., 471 U.S. 707, 713 (1985) (quoting Hines v. Davidowitz, 312 U.S. 52, 67 (1941)).

12.  *See generally* JEFFREY A. SEGAL & HAROLD J. SPAETH, THE SUPREME COURT AND THE ATTITUDINAL MODEL REVISITED (2002) (presenting the attitudinal model of judicial decision making, which predicts and explains judicial outcomes as products of judges' preferred social policies).

13.  *See, e.g.*, United States v. Lopez, 514 U.S. 549 (1995).

14.  *See* Frank B. Cross & Emerson H. Tiller, *The Three Faces of Federalism: An Empirical Assessment of Supreme Court Federalism Jurisprudence*, 73 S. CAL. L. REV. 741, 745 (2000); Michael S. Greve & Jonathan Klick, *Preemption in the Rehnquist Court: A Preliminary Empirical Assessment*, 14 SUP. CT. ECON. REV. 43, 79–80 (2006).

ernment (and against the states) is also a vote for 'big business' (and against pro-regulatory constituencies that want states to regulate above the federal baseline)."[15] Justices' political preferences stand in tension, making for "odd coalitions that appear to defy conventional left/right, liberal/conservative analysis."[16]

In response to this tension, says conventional wisdom, the conservative and liberal wings of the Court flip from their positions on federalism.[17] Conservatives can be expected to vote in favor of preemption and liberals to vote against it, with the odd result that the liberals find themselves promoting states' rights while conservatives counter with a plea for a robust national regulatory system.[18] Empirical evidence supports this conventional wisdom. Attitudinal models of judicial decision making rank the Rehnquist Court from most liberal to most conservative as follows: Justice Stevens anchors the liberal wing, followed by Justices Ginsburg, Breyer, Souter, O'Connor, Kennedy, Rehnquist, Scalia, and Thomas on the conservative pole.[19] The Rehnquist Court's voting record on preemption, with the exception of Justice Thomas, closely matches this array: Justice Stevens, on the liberal end, votes in favor of preemption in 41% of cases, while Justice Scalia, on the conservative end, votes in favor of preemption in 56% of cases, and the other Justices fall in line between.[20]

Yet, the conventional wisdom on judicial alignment only goes so far. When cases are considered in the aggregate, it is undoubtedly true that conservatives vote in favor of preemption more often than their liberal colleagues.[21] In any individual preemption case, however, both alignment and outcome are quite unpredictable.[22] Although alignments generally reflect conservative or liberal ideological leanings, it is always possible to pick up a vote from an "unlikely" justice.[23] No clear voting blocks have emerged to determine outcomes in close cases.[24] Thus, to give two particularly prominent examples, Justice Breyer sided with the Court's conservatives in favor of preemption in *Geier v. American Honda Motor Co.*,[25] and Justice Kennedy voted with the Court's liberals in *Wyeth v. Levine*.[26] Because preemption case outcomes depend on a large number of factors,[27] political ideology is, by itself, not a perfectly reliable indicator.[28] Alignments in preemption cases have an ideological tinge but remain somewhat fluid and unpredictable.

---

15.  Greve & Klick, *supra* note 14, at 79.
16.  Troy & Wood, *supra* note 6, at 260.
17.  *See* Greve & Klick, *supra* note 14, at 79–80.
18.  *See id.*
19.  *See* Theodore W. Ruger et al., *The Supreme Court Forecasting Project: Legal and Political Science Approaches to Predicting Supreme Court Decisionmaking*, 104 COLUM. L. REV. 1150, 1158 n.29 (2004).
20.  *See* Greve & Klick, *supra* note 14, at 81–82.
21.  *See infra* Table 1.
22.  *See* Greve & Klick, *supra* note 14, at 80 ("[P]reemption case law is not an exact mirror image of the Rehnquist Court's federalism: the voting alignments are substantially more fluid.").
23.  *Id.* at 83.
24.  *Id.* at 83–85.
25.  529 U.S. 861 (2000).

## IV. THE WYETH CLARIFICATION

Although ideological preferences play a significant role the Court's decision making, so too does the law itself, and federal preemption is an area where legal doctrine plays an especially significant part.[29] More so than most areas of the law, Justices tend to break from their traditional blocs, producing surprising alignments.[30] This fluidity renders preemption decisions unpredictable while also making the underlying preemption doctrines especially relevant to outcomes. Changes or clarifications in the preemption doctrine, therefore, deserve close attention. With its recent preemption decision in *Wyeth v. Levine*,[31] the Court has produced just such a clarification. In so doing, the Court has rendered outcomes more predictable in a certain category of preemption cases while also highlighting the role that legal doctrine should play in models of judicial decision making. This section analyzes the *Wyeth* decision, highlighting the specific doctrinal disagreements that separate the majority, concurrence, and dissent. These disagreements form the foundation for a later[32] discussion of the ways in which *Wyeth* clarifies the Court's conflict preemption jurisprudence.

### A. Factual Background and Legal Claims

In *Wyeth v. Levine*[33] the Court considered the common-law negligence claim of Diane Levine against the drug manufacturer Wyeth. Levine, suffering from a severe migraine headache, consented to a physician assistant's administration of Phenergan, a drug manufactured by the defendant.[34] The drug can be administered either intramuscularly or intravenously, and intravenous administration can be performed by either the slow IV-drip method or the faster but riskier

---

26. 129 S. Ct. 1187 (2009).
27. *See* Cross & Tiller, *supra* note 14, at 744–50; David B. Spence & Paula Murray, *The Law, Economics, and Politics of Federal Preemption Jurisprudence: A Quantitative Analysis*, 87 CALIF. L. REV. 1125, 1164–79 (1999).
28. *See* Carolyn Shapiro, *Coding Complexity: Bringing Law to the Empirical Analysis of the Supreme Court*, 60 HASTINGS L.J. 477, 487–88 (2009).
29. *See* Greve & Klick, *supra* note 14, at 79; Shapiro, *supra* note 24, at 487–88.
30. Troy & Wood, *supra* note 6, at 260.
31. 129 S. Ct. 1187 (2009).
32. *See infra* section IV.E and Parts V–VI.
33. 129 S. Ct. 1187 (2009).
34. *Id.* at 1191.

IV-push method.[35] Because her pain was severe and an initial administration of the drug had failed to provide relief, the physician assistant administered the drug via the IV-push method, which promises faster relief but also carries a risk of significant side effects.[36] The drug is corrosive, and if it escapes from the vein into surrounding tissue it causes irreversible gangrene.[37] Unfortunately, in Levine's case this precise danger was realized. As the physician assistant administered the drug, it escaped the vein and came in contact with arterial blood, resulting in gangrene and eventually requiring the amputation of Levine's right forearm.[38] As a result of this amputation, Levine incurred substantial medical expenses and was forced to abandon her career as a professional musician.[39]

Levine brought a common-law negligence action against Wyeth alleging that Phenergan was defectively labeled.[40] She argued that although the drug's label warned of the danger of gangrene following inadvertent intra-arterial injection, its labeling was nonetheless defective because it failed to instruct clinicians to use the IV-drip method as an alternative to the riskier IV-push method.[41] Wyeth responded by arguing that Levine's negligence claim was preempted by federal law. It urged the Court to find both impossibility and object preemption. First, it argued that it would have been impossible for it to comply with a state common-law duty to modify Phenergan's label while also remaining in compliance with Federal Drug Administration (FDA) regulations.[42] Second, it argued that recognition of the plaintiff's state tort action would create an unacceptable obstacle to the accomplishment of the purposes and objectives of Congress by substituting a lay jury's decision about drug labeling for the expert judgment of the FDA.[43]

## B. The Liberal Bloc's Majority Opinion

Justice Stevens wrote for the majority, joined by Justices Souter, Ginsburg, Breyer, and Kennedy.[44] The Court rejected both arguments. In response to Wyeth's impossibility preemption defense, the Court noted that although generally speaking a manufacturer may not change its label after it is approved by the FDA, federal regula-tions do provide a mechanism[45] for manufacturers to change a label to add to or strengthen warnings.[46] Because under federal law Wyeth would have been unilaterally able to strengthen the warnings on Phenergan's label, it was not impossible for Wyeth to comply with both federal and state requirements.[47]

---

35. *Id.*
36. *Id.*
37. *Id.*
38. *Id.*
39. *Id.*
40. *Id.* at 1191–92.
41. *Id.*
42. *Id.* at 1193–94.
43. *Id.*
44. *Id.* at 1190.

The Court also rejected Wyeth's obstacle preemption argument. Wyeth argued that Congress intended FDA regulations to establish both a floor and a ceiling for drug labeling requirements and that to allow a state negligence cause of action would interfere with that objective.[48] The Court, however, found no evidence of such an intent on the part of Congress. Looking to the legislative history of the Federal Food, Drug, and Cosmetic Act (FDCA), the Court noted that although the FDCA was intended to bolster consumer protection against harmful products, Congress intentionally provided no federal cause of action.[49] Instead, Congress decided to rely on state tort law to provide appropriate relief.[50] Such a policy is inconsistent with the notion that Congress intended to preempt state common law. Furthermore, reasoning from congressional silence, the Court noted that if Congress thought that state tort law posed an obstacle to the objectives of the FDCA, "it surely would have enacted an express pre-emption provision at some point during the FDCA's 70-year history."[51] Thus, concluding that Congress did not intend to displace state law, the Court reasoned that state tort actions posed no obstacle to the achievement of congressional purposes and objectives.

---

45. FDA regulations permit drug manufacturers to modify labels "[t]o add or strengthen a contraindication, warning, precaution, or adverse reaction . . . [or to] add or strengthen an instruction about dosage and administration that is intended to increase the safe use of the drug product." 21 C.F.R § 314.70(c)(6)(iii)(A), (C) (2010).

46. *Wyeth*, 129 S. Ct. at 1196.

47. *Id.* at 1199.

48. *Id.*

49. *Id.* at 1199–1200 & n.7 (discussing House Resolution 6110, 73d Cong. § 25 (1933), and citing testimony that no federal cause of action is necessary because common-law claims were already available under state law).

50. *Id.* at 1200.

51. *Id.* ("[Congress's] silence on the issue [of preemption], coupled with its certain awareness of the prevalence of state tort litigation, is powerful evidence that Congress did not intend FDA oversight to be the exclusive means of ensuring drug safety and effectiveness."). The Court also noted Congress's 1976 enactment of an express pre-emption provision for medical devices. Congress could have applied this preemption provision to the FDCA in its entirety, including prescription drug requirements, but instead chose to apply the clause only to medical devices. *Id.*

## C. Justice Thomas's Concurrence

Justice Thomas, concurring only in judgment, agreed with the majority that FDA labeling requirements did not preempt Levine's common-law negligence claim, but Justice Thomas's rationale was markedly different.[52] Rather than analyzing Wyeth's obstacle preemption argument under the Court's long-standing framework, Justice Thomas boldly rejected as unsound the Court's entire line of obstacle preemption jurisprudence.[53] Drawing on the theory of dual sovereignty embodied in Federalist No. 51,[54] as well as the Bicameral and Presentment Clause requirements of the Constitution,[55] Thomas emphasized his "increasing[ ] reluctan[ce] to expand federal statutes beyond their terms through doctrines of implied pre-emption."[56] Preemption, he argued, must turn on something more than "generalized notions of congressional purposes that are not contained within the text of federal law."[57] Preemption analysis of this sort is a "potentially boundless" intrusion on state sovereignty by the federal judiciary without congressional authorization.[58] Preemption must instead be grounded in a direct conflict between state law and the text of validly promulgated federal law.[59]

Perhaps recognizing that the complete elimination of obstacle preemption could have radical and possibly undesirable consequences, Justice Thomas also urged the Court to adopt a slightly broader view of impossibility preemption.[60] Though he offers few details, Justice Thomas hints at what a broader impossibility doctrine might look like. He notes that the Supremacy Clause does not limit direct conflicts to cases of literal physical impossibility,[61] and he instead suggests that the analysis should focus on whether state law logically contradicts or is repugnant to federal law.[62] However, even under this broader impossibility analysis—whatever its precise contours—Justice Thomas concluded that the text of the federal laws at issue did not preempt the plaintiff's state law claim. Federal law provides a mechanism for manufacturers to strengthen drug labels, and so it would have been entirely possible for Wyeth to update Phenergan's label to account for the dangers of IV-push administration.[63]

---

52. *Id.* at 1204–05 (Thomas, J., concurring in judgment).

53. *Id.* at 1211 ("This Court's entire body of 'purposes and objectives' pre-emption is inherently flawed. The cases improperly rely on legislative history, broad atextual notions of congressional purpose, and even congressional inaction in order to pre-empt state law.").

54. *Id.* at 1205 (quoting THE FEDERALIST NO. 51, at 266 (James Madison) (M. Beloff ed., 2d ed. 1987)).

55. U.S. CONST. art. I, § 7, cl. 2 ("Every Bill which shall have passed the House of Representatives and the Senate, shall, before it become a Law, be presented to the President of the United States . . . .").

56. *Wyeth*, 129 S. Ct. at 1207 (quoting Bates v. Dow Agrosciences LLC, 544 U.S. 431, 459 (2005) (Thomas, J., concurring in judgment)).

57. *Id.*

58. *See id.* at 1207 (quoting Geier v. Am. Honda Motor Co., 529 U.S. 861, 907 (2000) (Stevens, J., dissenting) (internal quotation marks omitted)).

59. *Id.* at 1208.

60. *Id.* at 1209 ("The Court has generally articulated a very narrow 'impossibility standard'—in part because the overly broad sweep of the Court's 'purposes and objectives' approach has rendered it unnecessary for the Court to rely on 'impossibility' pre-emption. . . . Therefore 'physical impossibility' may not be the most appropriate standard for determining whether the text of state and federal laws directly conflict.") (citations omitted).

### D. The Conservative Bloc's Dissent

Unlike Justice Thomas, Justice Alito, joined by Justice Scalia and Chief Justice Roberts in dissent,[64] expressed no hesitation whatsoever in applying the Court's traditional obstacle preemption framework. Indeed, he took the opposite position, criticizing the majority for an overly narrow view of obstacle preemption.[65] "Congress made its 'purpose' plain in authorizing the FDA—not state tort juries—to determine when and under what circumstances a drug is 'safe,'"[66] and by finding unpreempted the plaintiff's state tort law claims, the majority "upset the regulatory balance struck by the federal agency."[67] Justice Alito argued that state common law is fundamentally inconsistent with the federal objective of providing standardized safety determinations and that state law must yield in the face of such inconsistency.[68]

---

61. Professor Nelson provides a useful example of the somewhat broader reach of a logical contradiction test. Imagine that a federal law gives workers the right to join labor unions but that a state law forbids unionization or holds union members liable for damages if they choose to unionize. It would be physically possible for an individual to comply with both federal and state law. All he would need to do is refrain from joining a union. Nonetheless, there is a logical contradiction between the two laws that a court presented with the issue would need to address. By enforcing the state law a court would be ignoring the federal right. *See* Nelson, *supra* note 5, at 260–61.

62. *Wyeth*, 129 S. Ct. at 1209. Justice Thomas references both Professor Nelson's language of "logical-contradiction" and Justice Story's language of constitutional repugnancy, but it is unclear whether he views these two standards as coterminous, whether he prefers one construct over the other, or whether he might have in mind a standard with an even broader scope that simply takes logical contradiction and repugnancy as a starting point. *Id.* (citing 3 JOSEPH STORY, COMMENTARIES ON THE CONSTITUTION § 1836 (Boston, Hilliard, Gray & Co. 1833); Nelson, *supra* note 5, at 260–61).

63. *Id.* at 1210.

64. *Id.* at 1217 (Alito, J., dissenting).

65. *Id.* at 1220 ("A faithful application of this Court's conflict pre-emption cases compels the conclusion that the FDA's 40-year-long effort to regulate the safety and efficacy of Phenergan pre-empts respondent's tort suit. Indeed, that result follows directly from our conclusion in *Geier*.").

66. *Id.* at 1219.

67. *Id.* at 1220.

68. *Id.* at 1221.

## E.  Lessons and Unanswered Questions

*Wyeth* is, of course, only a single case, and it is important to avoid overemphasizing its significance.  Nonetheless, a careful reading of *Wyeth* and the Court's prior conflict preemption decisions yields interesting results.  Note the alignment of the Justices.  In favor of preemption are, as the attitudinal model would suggest,[69] Chief Justice Roberts and Justices Scalia and Alito of the Court's conservative core.  Against preemption are, again as the attitudinal model would suggest, Justices Stevens, Souter, Ginsburg, and Breyer.  Although Justice Kennedy's vote against preemption might initially seem surprising, he is a moderate conservative,[70] falling roughly in the center of the Court's ideological spectrum,[71] so it is not entirely unexpected that he would vote unpredictably in any particular case.

Only Justice Thomas appears out of place.  As a conservative, one might expect him to vote in favor of preemption, and in a great number of cases this intuition would be accurate.[72]  In obstacle preemption cases, however, Justice Thomas's voting history shows a sharp movement leftward, disfavoring preemption in the absence of an express preemption clause.[73]  *Wyeth* represents the culmination of that leftward movement.  After years of votes evincing his dissatisfaction with the Court's sometimes broad obstacle preemption doctrine,[74] Justice Thomas in *Wyeth* declares his complete abandonment of that preemption rationale, marking himself as a reliable vote against obstacle preemption in any future decision.

Justice Thomas's rejection of obstacle preemption, however, does not by itself shed much light on the Court's preemption jurisprudence.  The idiosyncratic views of a single Justice are not likely to sway a majority in future cases.[75]  What makes Justice Thomas's position important is the identity (and quantity) of those Justices with whom he joins in concurrence.  Complementing Justice Thomas's complete re-jection of obstacle preemption in *Wyeth* was a group of Justices who reliably construe purposes and objectives preemption narrowly, reserving the doctrine for only the most direct conflicts between state law and congressional objectives—namely, Justices Ginsburg, Stevens, Souter, and Breyer. If Justice Thomas is, along with the Court's liberal wing, a reliable vote against obstacle preemption, we might expect to see a predictable bloc of five Justices countering and trumping the preemption votes of the Court's conservative wing.

---

69.  *See supra* Part III.

70.  *Id.*

71.  Justice Kennedy, is, however, markedly more conservative on the issue of preemption than in his jurisprudence generally.  He is firmly within the conservative camp on preemption questions.  *See infra* Part V.

72.  *See, e.g.*, Altria Grp., Inc. v. Good, 555 U.S. 70 (2008); Rowe v. N.H. Motor Trans. Ass'n, 552 U.S. 364 (2008); Riegel v. Medtronic, Inc., 552 U.S. 312 (2008); Merril Lynch v. Dabit, 547 U.S. 71 (2006); Bates v. Dow Agrosciences LLC, 544 U.S. 431 (2005).

73.  *See infra* Part V.

74.  *See* Wyeth v. Levine, 129 S. Ct. 1187, 1207 (2009) (Thomas, J., concurring in judgment) (collecting cases).

75.  Consider Justice Thomas's idiosyncratic view that the Federal Arbitration Act should not be read to preempt state law, which, despite repeated lone dissents, has failed to attract followers.  *See, e.g.*, Doctor's Assocs., Inc. v. Casarotto, 517 U.S. 681 (1996); Mastrobuono v. Shearson Lehman Hutton, Inc., 514 U.S. 52 (1995).

This feature of the Court's alignment in *Wyeth* raises two important sets of questions. First, exactly how cohesive is this five-Justice bloc against obstacle preemption? Justice Thomas's vision of expanded impossibility preemption sounds very much like the liberal bloc's historically narrow version of obstacle preemption.[76] Are they essentially the same analysis under different names, and, if so, do we now have five Justices determining outcomes under the same framework? Second, if Justice Thomas is engaging in approximately the same inquiry as the Court's liberal wing, what can be learned from that? What factors does this emerging five-Justice bloc find most relevant to the obstacle preemption inquiry?

Fortunately, both sets of questions are susceptible to empirical investigation. The following sections take up each question in turn. First, Part V identifies and examines every obstacle preemption case from the stabilization of the Court's conservative–liberal alignment at the start of the "Second" Rehnquist Court in the 1994–95 Term until the present. Justice Thomas's decision to abandon obstacle preemption was the result of a gradual drift, and so, presumably, he has been voting more or less consistently with his newly announced view for some time. A review of the Court's obstacle preemption decisions over the last fifteen years reveals both how closely his view parallels that of the Court's liberal wing and how cohesively the Court's liberal wing itself votes in obstacle preemption cases. Second, Part VI builds on Part V's study of the Court's prior decisions to identify the factors that the loosely aligned five-Justice bloc finds most relevant in deciding obstacle preemption cases.

## V.  AN EMPIRICAL STUDY OF OBSTACLE PREEMPTION IN THE SUPREME COURT FROM 1994–2009

A search of the Westlaw "Supreme Court Cases" database for all cases after 1993[77] containing terms or phrases indicative of obstacle preemption[78] returned 128 results. This search was, by design, over-inclusive, and a closer examination revealed that only twenty-five of the cases involved a genuine question of obstacle preemption. For each case an entry was created in a database recording the outcome of the case as well as the vote of each individual Justice for or against obstacle preemption. In a few cases it was impossible to determine how a particular Justice would have voted on the issue of obstacle preemption, because that Justice's reasoning decided the case on an alternative ground. For example, in *Mid-Con Freight Systems, Inc. v. Michigan Public Service Commission*,[79] a six-Justice majority clearly sided against obstacle preemption, but three dissenting Justices would have found express preemption[80] and accordingly failed to reach the question of obstacle preemption. In such cases, clear votes for or against obstacle preemption were included in the database while ambiguous votes were omitted.

---

76.  Hines v. Davidowitz, 312 U.S. 52, 67 (1941).
77.  Justice Breyer was appointed to the Supreme Court in 1994.
78.  The precise search string was as follows: "pre-empt! & da(aft 1993) & ('obstacle' 'impossib!' 'implied' 'conflict' 'purposes and objectives')".

## A.   General Observations

From a high-level perspective the Court's obstacle preemption decisions look much like its express, field, and impossibility preemption decisions. Over the period studied, obstacle preemption was found in 50% of cases, compared to 52% in all types of preemption cases considered collectively.[81] The percentage of obstacle preemption cases decided unanimously also appears relatively consistent with all other types of preemption. Fifty percent of obstacle preemption cases were decided unanimously with an additional 18% contested only lightly by two or fewer Justices. This degree of unanimity is somewhat surprising given the Court's 40% unanimity rate for all cases, but it is consistent with the Court's higher-than-average unanimity rate of 51% in preemption cases generally.[82] These data show that preemption is slightly less divisive than other issues to come before the Court and that, if anything, the Court is somewhat less likely to find obstacle preemption than express or field preemption.

## B.   Ideological Realignment on Obstacle Preemption

Although the Court's obstacle preemption decisions appear almost identical to its general preemption decisions when considered from a high level, a closer examination reveals surprising changes in the Court's alignment. Consider Table 1 and Figure A below, which compare the probability of a given Justice voting in favor of obstacle preemption to the probability of that Justice voting for preemption generally. Although the Court as a whole is about as likely to find obstacle preemption as preemption generally, the individual Justices vote quite differently depending on the type of preemption involved.

---

79.  545 U.S. 440 (2005).
80.  *Id.* at 464.
81.  Greve & Klick, *supra* note 14, at 57.
82.  *Id.* at 56.

The data indicate movement in both directions along the ideological spectrum.  Chief Justice Rehnquist and Justices Kennedy and Stevens are all, perhaps surprisingly,[83] somewhat more likely to find obstacle preemption than other varieties.  Justices Thomas and Ginsburg, on the other hand, move in the opposite direction, voting for obstacle preemption less frequently than for preemption generally.

| Justice | Obstacle Preemption | Preemption Generally[84] |
|---------|---------------------|--------------------------|
| Kennedy | .61 | .53 |
| Scalia | .57 | .56 |
| Rehnquist | .58 | .50 |
| O'Connor | .53 | .52 |
| Stevens | .48 | .41 |
| Souter | .46 | .43 |
| Breyer | .44 | .52 |
| Ginsburg | .38 | .41 |
| Thomas | .35 | .44 |
| Court | .50 | .52 |

Table 1.    Probability of Voting in Favor of Obstacle Preemption vs. Preemption Generally

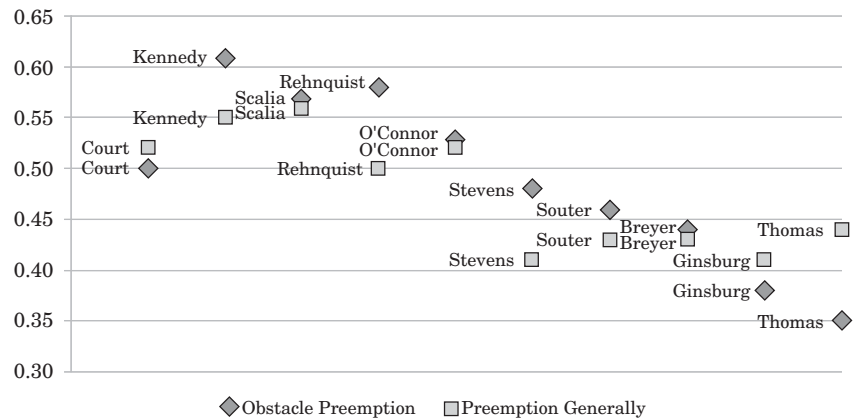

Figure A

---

The data appear to confirm Justice Thomas's alignment with the liberal bloc on the issue of obstacle preemption. Unlike preemption generally, where he retains his conservative pro-preemption stance, on questions of obstacle preemption Justice Thomas is firmly in the liberal anti-preemption camp, voting for obstacle preemption only 35% of the time—less than any other Justice.[85] Other significant shifts along the ideological spectrum are apparent as well. These movements can be seen clearly in Table 2 below, which compares the ideological rankings of each Court member in all cases generally, in all preemption cases, and in obstacle preemption cases specifically. Rankings are according to the percentage of cases in which each Justice has voted in what is generally regarded as a liberal or conservative fashion, with one being the most and nine the least conservative. Justice Thomas's shift is the most severe. He is generally regarded to be the Court's most conservative member,[86] but in preemption cases he usually shifts to the center, and in obstacle preemption cases specifically he shifts all the way to the leftmost position on the Court. Justice Kennedy also shifts substantially, but in the opposite direction. Although he is regarded as a centrist,[87] falling squarely in middle of the Court ideologically, in preemption cases he moves decidedly rightward, voting in favor of obstacle preemption more frequently than any other Justice. We see a second movement rightward in Justice Stevens who, though often regarded as the Court's most liberal

| Justice | General Attitudinal Model of All Cases | All Preemption | Obstacle Preemption |
|---------|:---:|:---:|:---:|
| Thomas | 1 | 5 | 9 |
| Scalia | 2 | 1 | 3 |
| Rehnquist | 3 | 4 | 2 |
| Kennedy | 4 | 2 | 1 |
| O'Connor | 5 | 3 | 4 |
| Souter | 6 | 6 | 6 |
| Breyer | 7 | 6 | 7 |
| Ginsburg | 8 | 8 | 8 |
| Stevens | 9 | 8 | 5 |

Table 2.    Ideological Rank by Case Type

85. Given Justice Thomas's complete renunciation of obstacle preemption doctrine in *Wyeth*, it's not surprising that his votes in favor of it are few and far between. Indeed, one might question how it could be that he *ever* votes for obstacle preemption. The answer may lie, to some extent, in his unique vision of impossibility preemption. *See infra* section V.A.

86. Robert H. Smith, *Uncoupling the "Centrist Bloc": An Empirical Analysis of the Thesis of a Dominant, Moderate Bloc on the United States Supreme Court*, 62 Tenn. L. Rev. 1, 1 (1994).

87. *Id.* at 1–2.

member,[88] shifts rightward into a centrist position on questions of obstacle preemption.

These movements may at first glance seem relatively insignificant. A few Justices are slightly more likely to vote for obstacle preemption than others, but for most, obstacle preemption is about a fifty–fifty proposition, and, as shown by Table 1 above, the case outcomes reflect that. This characterization, though, would be misleading. In an area of the law where 50% of cases are decided unanimously, it's unsurprising that most Justices' decision rates would be close to those of the Court as a whole. In closely contested cases, however, in which at least three Justices dissent from the majority view, the Court is substantially more polarized. In such cases the law points clearly in neither direction, and the Justices' differences are magnified, as reflected in Table 3 and Figure B. Chief Justice Rehnquist and Justices O'Connor, Stevens, and Breyer diverge only slightly from their averages in all obstacle preemption cases. Justices Kennedy and Scalia favor preemption more strongly, and Justices Souter, Thomas, and Ginsburg more strongly disfavor it.

| Justice | Contested Obstacle Preemption Cases | All Obstacle Preemption Cases |
|---------|-------------------------------------|-------------------------------|
| Kennedy | .71 | .61 |
| Scalia | .71 | .57 |
| Rehnquist | .60 | .58 |
| O'Connor | .60 | .53 |
| Stevens | .43 | .48 |
| Souter | .29 | .46 |
| Breyer | .43 | .44 |
| Ginsburg | .14 | .38 |
| Thomas | .29 | .35 |
| Court | .43 | .50 |

Table 3.   Probability of Voting for Obstacle Preemption

The comparison between the Court's obstacle preemption and general preemption decisions suggests important positional shifts among the Justices based on the case's theory of preemption. Justice Thomas is a moderate on preemption generally but strongly disfavors obstacle preemption. Justice Stevens, on the other hand, shifts in the opposite direction, disfavoring obstacle preemption less strongly than preemption generally. This suggests that models of judicial decision making that fail to account for legal doctrine may, at least in the context of preemption, be ignoring an important factor in the Courts decisions.[89]

---

88. *Id.* at 1.

89. *Cf.* Shapiro, *supra* note 28, at 487 ("There can, of course, be no question that policy preferences or ideology play a role in Supreme Court decision making. But

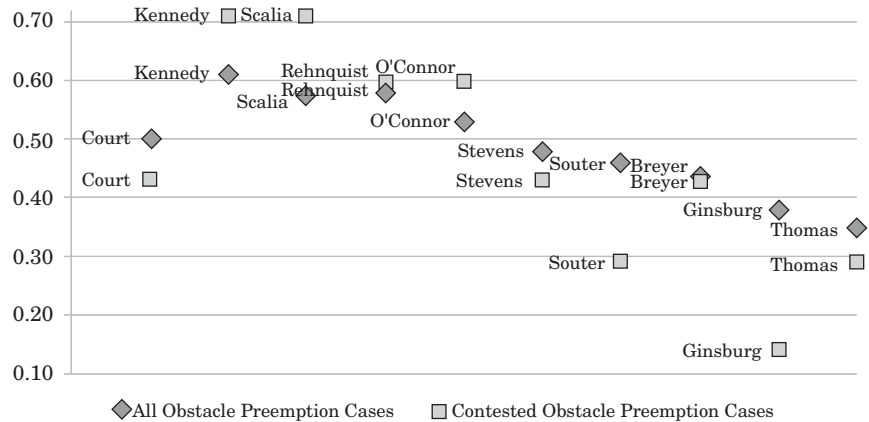

Figure B

Court watchers and potential litigants would be well-advised to consider the shifts in judicial outlook that occur in cases of obstacle preemption.

## C. Obstacle Preemption Voting Blocs

The data clearly confirm *Wyeth*'s suggestion[90] of an ideological reshuffling of the Court in the context of obstacle preemption to an alignment in which five Justices can generally be expected to vote against obstacle preemption, and four Justices can generally be expected to vote in favor of it. Yet, this does not alone imply the existence of a consistent voting bloc against obstacle preemption. It is entirely possible that, although individual Justices exhibit ideological preferences in one direction or another, they do not regularly vote together in a consistent bloc. Indeed, this is precisely what occurs in preemption cases generally: "Conservative justices tend to vote for preemption in many cases, and liberal justices tend to do the opposite. But neither side seems to agree on what cases, precisely, call for the 'default' response."[91]

In the context of obstacle preemption, however, the Court appears to divide into somewhat more cohesive groups. Five Justices cannot only be expected to vote against obstacle preemption in the majority of

---

the interesting and important contemporary questions—and the ones that cannot be answered using [the Spaeth Database]—are whether and how much law matters as well, how ideology and law interact with or affect each other, and how these interactions vary from case to case or from Justice to Justice.").

90. *See supra* Part IV.
91. Greve & Klick, *supra* note 14, at 84.

cases, but they can be expected to vote together against obstacle preemption in the same cases. Consider Table 4 below, which shows the frequency with which each Justice has voted with the majority in contested obstacle preemption decisions. The data show no absolute voting bloc. Rather, they show a semicohesive bloc that, while not absolute, is determinative in the majority of cases. Note, for instance, that Justice Breyer has formed part of the majority in every contested obstacle preemption decision. Justice Souter is not far behind, forming part of the majority in all but one case.

| Justice | Pro-Preemption (3) | Anti-Preemption (4) | All Contested (7) |
|---------|--------------------|--------------------|-------------------|
| Kennedy | 3 | 2 | 5 |
| Scalia | 2 | 1 | 3 |
| Rehnquist | 1 | 1 | 2 |
| O'Connor | 1 | 1 | 2 |
| Stevens | 2 | 3 | 5 |
| Breyer | 3 | 4 | 7 |
| Souter | 2 | 4 | 6 |
| Thomas | 1 | 3 | 4 |
| Ginsburg | 1 | 4 | 5 |

Table 4.   Frequency of Voting with the Majority in Contested Holdings

This anti-obstacle preemption bloc can be seen somewhat more clearly by looking at the probability of individual Justices voting together in contested obstacle preemption cases, as exhibited by Table 5 below. It is the Court's liberal wing and Justice Thomas that most often vote with the majority. Note, too, that the strongest alliance pairs exist among these Justices. Justice Souter votes with Justices Stevens, Breyer, and Ginsburg 86% of the time and Justice Thomas 71% of the time. The data show no comparable conservative alliances. Still, it is important not to overemphasize the significance of these alliances. They are far from absolute. Justice Thomas votes with Justice Scalia as often as with Justices Stevens and Breyer and with the Court as a whole only slightly more than half of the time. The data

| Justice | Kennedy | Scalia | Rehnquist | O'Connor | Stevens | Breyer | Souter | Thomas | Ginsburg |
|---------|---------|--------|-----------|----------|---------|--------|--------|--------|----------|
| Kennedy | 1.00 | | | | | | | | |
| Scalia | .43 | 1.00 | | | | | | | |
| Rehnquist | .80 | .40 | 1.00 | | | | | | |
| O'Connor | .40 | .40 | .60 | 1.00 | | | | | |
| Stevens | .43 | .43 | 0.00 | .40 | 1.00 | | | | |
| Breyer | .71 | .43 | .40 | .40 | .71 | 1.00 | | | |
| Souter | .57 | .29 | .20 | .20 | .86 | .86 | 1.00 | | |
| Thomas | .57 | .57 | .40 | 0.00 | .57 | .57 | .71 | 1.00 | |
| Ginsburg | .43 | .14 | .40 | .40 | .71 | .71 | .86 | .57 | 1.00 |
| Court | .71 | .43 | .40 | .40 | .71 | 1.00 | .86 | .57 | .71 |

Table 5.   Justices' Probability of Voting Together in Contested Obstacle Preemption Cases

show a weak but nonetheless significant voting bloc against obstacle preemption.

The data also appear to disconfirm the hypothesis that Justice Thomas's newly announced, expanded view of impossibility preemption may be essentially identical to the liberal wing's narrow view of obstacle preemption.[92] Although he did not formally declare his complete rejection of obstacle preemption and his corresponding embrace of broader impossibility preemption until *Wyeth*, Justice Thomas has been voting more or less in accord with these views for some time.[93] If his expanded impossibility preemption were identical to the liberal wing's narrow preemption, we would see a much higher correlation between his votes and the rest of the anti-obstacle preemption bloc. Instead, despite being the Court's most reliable obstacle preemption opponent, he appears to be the outlier of the anti-preemption coalition. This suggests that the bloc's votes are the products of different frameworks—that although Justice Thomas's broadened impossibility preemption and the liberals' narrow obstacle preemption analyses often produce the same non-preemptive outcomes, their differing underlying premises occasionally produce divergent results.

## VI.   COUNTING TO FIVE: THE COURT'S EMERGING CONFLICT PREEMPTION ANALYSIS

The previous section showed a weak anti-obstacle preemption bloc that more often than not votes together but sometimes fractures in contested cases. It is from different doctrinal routes that Justice Thomas and the rest of the bloc come to be aligned, so it is not overly surprising that the alliance is not ironclad. This section highlights the differences between Justice Thomas's approach and that of the rest of the bloc, enabling court watchers and potential litigants to predict outcomes and present their cases more effectively.

---

92. *See* Wyeth v. Levine, 129 S. Ct. 1187, 1209 (Thomas, J., concurring in judgment) ("The Court has generally articulated a very narrow 'impossibility standard'—in part because the overly broad sweep of the Court's 'purposes and objectives' approach has rendered it unnecessary for the Court to rely on 'impossibility' preemption. . . . Therefore, 'physical impossibility' may not be the most appropriate standard for determining whether the text of state and federal laws directly conflict") (citations omitted); *supra* section III.E.

93. Indeed, he suggests as much in *Wyeth*: "I have become 'increasing[ly] reluctan[t] to expand federal statutes beyond their terms through doctrines of implied preemption.' My review of this Court's broad implied pre-emption precedents, particularly its 'purposes and objectives' pre-emption jurisprudence, has increased my concerns that implied pre-emption doctrines have not always been constitutionally applied." *Wyeth*, 129 S. Ct. at 1207 (quoting Bates v. Dow Agrosciences LLC, 544 U.S. 431, 459 (2005) (Thomas, J., concurring in judgment); Pharm. Research & Mfrs. of Am. v. Walsh, 538 U.S. 644, 678 (2003) (Thomas, J., concurring in judgment)).

## A. Justice Thomas

As seen most recently in his *Wyeth* concurrence, the key factor in Justice Thomas's preemption analysis is the explicitness of congressional action. Absent clear action by Congress to preempt state law, states should be presumed to retain their sovereignty.[94] Any other judicial approach would aggrandize the judiciary at the expense of the legislature and violate the principle of dual sovereignty enshrined in the Constitution.[95] Justice Thomas appears to take quite seriously (more so than the rest of the Court) the Court's oft-repeated rules that "the purpose of Congress is the ultimate touchstone in every preemption case"[96] and that "we start with the assumption that the historic police powers of the States were not to be superseded by the Federal Act unless that was the clear and manifest purpose of Congress."[97]

This concern for state sovereignty and the separation of powers affects even Justice Thomas's express preemption jurisprudence. In *CSX Transportation, Inc. v. Easterwood*,[98] for example, he, joined by Justice Souter, dissented from the majority's interpretation of the Federal Railroad Safety Act (FRSA) to preempt a common-law negligence claim by the widow of a man killed at a railroad crossing.[99] The majority reasoned that the FRSA's express preemption of conflicting state laws precluded the plaintiff's claim that the train was traveling at an unsafe speed, because regulations enacted pursuant to the FRSA already set maximum speeds based on track characteristics.[100] Justice Thomas disagreed and would have required more explicit regulatory action before finding preemption. He reasoned that regulations setting speed maximums with regard to track characteristics could not reasonably be interpreted as preempting state law regulating safe train speed through grade crossings.[101]

Even in his newly expanded view of impossibility preemption, where congressional preemptive intent must be inferred, Justice Thomas is likely to insist quite rigidly on readily-discernible intent. The Court's finding of preemption in *Barnett Bank of Marion County v. Nelson*[102] provides an example of what Justice Thomas's new impossibility preemption might look like. In that case the Court found preempted a Florida state law prohibiting national banks from selling insurance, because it interfered with Congress's purposes and objec-tives in enacting the McCarran–Ferguson Act of 1916, which permits national banks to sell insurance in small towns.[103] Post-*Wyeth*, the obstacle preemption doctrine would, of course, be unavailable to Justice Thomas. The *Nelson* case, however, fits neatly within an expanded impossibility framework. Compliance with both state and federal law is not literally impossible. National banks could simply refrain from selling insurance in small Florida towns, despite the fact that federal law appears explicitly to permit it. Nonetheless the state and federal laws are impossible to reconcile in a less literal yet still very palpable sense[104]—a sense that Justice Thomas would rely on as indicating congressional preemptive intent despite the lack of an express preemption clause.

---

94.  Wyeth v. Levine, 129 S. Ct. 1187, 1205–07 (Thomas, J., concurring in judgment).
95.  *See Wyeth*, 129 S. Ct. at 1205–07.
96.  Medtronic, Inc. v. Lohr, 518 U.S. 470, 485 (1996) (internal quotation marks omitted).
97.  Rice v. Santa Fe Elevator Corp., 331 U.S. 218, 230 (1947).
98.  507 U.S. 658 (1993).
99.  *Id.* at 676.
100.  *Id.* at 673–76.
101.  *Id.* at 677 (Thomas, J., concurring in part and dissenting in part).
102.  517 U.S. 25 (1996).

In one category of cases, however, Justice Thomas appears willing to settle for something slightly less than perfect evidence of congressional intent. Where an express preemption clause is clearly intended to preempt something, but that something is not entirely clear, Justice Thomas appears willing to invoke principles of obstacle preemption to resolve the question of preemptive scope. For instance, in *Rush Prudential HMO, Inc. v. Moran*,[105] the Court considered a state insurance law that Rush Prudential argued was preempted by the Employee Retirement Income Security Act.[106] The majority found the state provision to fall within the Employee Retirement Income Security Act's savings clause, which exempts state laws regulating insurance, banking, or securities.[107] Justice Thomas, however, joined by Chief Justice Rehnquist and Justices Scalia and Kennedy,[108] reasoned that even a law regulating insurance could be preempted under "ordinary principles of conflict pre-emption" if it stood as a sufficient obstacle to Congress's broader purposes in enacting the statute.[109] Thus, it appears that Justice Thomas's deference to state sovereignty is lessened where Congress has already evidenced at least some intent to preempt and the primary question is one of preemptive scope.

## B. The Liberal Wing

Like Justice Thomas, Justices Souter, Stevens, Breyer, and Ginsburg side against obstacle preemption on a fairly consistent basis. However, to the extent that it is possible to generalize about all four collectively, they appear to do so for markedly different reasons. Whereas Justice Thomas's position stems almost entirely from his re-spect for the constitutional principles of dual sovereignty and separation of powers, the Court's liberal wing deploys a wider variety of arguments in support of its limited obstacle preemption jurisprudence. Unlike Justice Thomas, for instance, who reserves the presumption against preemption for use only in determining whether Congress intended federal law to have preemptive effect,[110] the other members of the anti-obstacle preemption bloc deploy the presumption more broadly to determine the preemptive scope of federal law as well as its preemptive effect.[111] The presumption serves as a thumb on the scale against preemption throughout the entire preemption analysis.

---

103.   *Id.* at 28–29.
104.   *See* Nelson, *supra* note 5, at 260–61.
105.   536 U.S. 355 (2002).
106.   *Id.* at 366–67.
107.   *Id.* at 387.
108.   *Id.* at 388.
109.   *See id.* at 394 (Thomas, J., dissenting).

Similarly, these Justices rely on a second interpretive principle eschewed by Justice Thomas: unless its intent to do so is especially clear, Congress should not be presumed to preempt state law where federal preemption would remove all means of judicial recourse.[112] The principle has intuitive appeal, but, as Justice Thomas notes, it often relies on inferences from congressional silence or ambiguous legislative history.[113] Interestingly, however, Justice Thomas's rejection of obstacle preemption has a similar function. He would require explicit congressional intent not only where preemption would remove all judicial recourse, but in all cases, whether or not an alternative remedy exists. Thus, the interpretive principle is a sort of middle ground that allows those Justices who use it to fine-tune preemption doctrine to suit the facts of particular cases, while Justice Thomas relies on a blunter instrument.

---

110. *See* Cipollone v. Liggett Grp., Inc., 505 U.S. 504, 544–45 (1992) (Scalia, J., dissenting).

111. *See* Wyeth v. Levine, 129 S. Ct. 1187, 1195 n.3 (2009) ("We rely on the presumption because respect for the States as independent sovereigns in our federal system leads us to assume that Congress does not cavalierly pre-empt state-law causes of action.") (quoting Medtronic, Inc. v. Lohr, 518 U.S. 470, 485 (1996)) (internal quotation marks omitted); Bates v. Dow Agrosciences LLC, 544 U.S. 431, 449 (2005) ("Because the States are independent sovereigns in our federal system, we have long presumed that Congress does not cavalierly pre-empt state-law causes of action.") (quoting Medtronic, Inc. v. Lohr, 518 U.S. 470, 485 (1996)) (internal quotation marks omitted); Rice v. Santa Fe Elevator Corp., 331 U.S. 218, 230 (1947).

112. *See Bates*, 544 U.S. at 449–50 ("The long history of tort litigation against manufacturers of poisonous substances adds force to the basic presumption against pre-emption. If Congress had intended to deprive injured parties of a long available form of compensation, it surely would have expressed that intent more clearly.") (citing Silkwood v. Kerr-McGee Corp., 464 U.S. 238, 251 (1984)); Riegel v. Medtronic, 552 U.S. 312, 1016 (2008) (Ginsburg, J., dissenting) ("It is difficult to believe that Congress would, without comment, remove all means of judicial recourse for large numbers of consumers injured by defective medical devices.") (quoting Silkwood v. Kerr-McGee Corp., 464 U.S. 238, 251 (1984)) (internal quotation marks omitted).

113. *Compare Bates*, 544 U.S. at 449–50 (Justice Stevens, joined by Chief Justice Rehnquist and Justices O'Connor, Kennedy, Souter, Ginsburg, and Breyer, relying on the presumption), *with id.* at 456–58 (Thomas, J., concurring in part and dissenting in part) (rejecting the majority's reliance on the historical availability of a state-law remedy as an improper inference from silence in the legislative history).

Finally, the liberal wing, unlike Justice Thomas, is quite willing to rely on legislative history to infer broad notions of congressional purpose.[114] Indeed, this is, in a sense, the crux of the entire debate between Justice Thomas and the rest of the Court. Should the Court defer to an imprecise notion of congressional purpose (whether from legislative history or elsewhere), or should the Court set a more rigorous standard of preemptive intent? This willingness to look to legislative history and congressional intent may explain why it is Justice Thomas and not one of the other anti-obstacle preemption bloc members who has the most consistent record of votes against obstacle preemption.[115] Without legislative history, he has no reason upon which to ground a vote for preemption, even if that result appears desirable in a particular instance. The other bloc members, on the other hand, have more freedom to find for preemption where they deem circumstances to require it. Thus, for instance, it is Justice Breyer who jumps ship to favor preemption in *Geier v. American Honda Motor Co.*,[116] citing legislative history to buttress his argument that the imposition of additional common-law vehicle safety requirements would interfere with Congress's purposes and objectives in passing the National Traffic and Motor Vehicle Safety Act.[117] Justice Thomas, not at all hesitant to find preemption when expressly mandated by statute,[118] remained firmly opposed, citing the Act's savings clause as reason to curtail significantly its preemptive scope.[119]

Collectively these factors show a divide between Justice Thomas and the rest of the anti-obstacle preemption bloc. While Justice Thomas's obstacle preemption jurisprudence is heavily dependant on a single factor—explicit congressional authorization—the other Justices employ a more varied arsenal, considering such factors as legislative history and intent, availability of alternative remedies, and the presumption against infringing on areas of traditional state sovereignty.[120] As a result Justice Thomas, limited to a single determini-tive factor, rarely sides in favor of obstacle preemption, while the other members of the bloc can be expected to vote in favor of obstacle preemption any particular case depending on the significance they place on each relevant decisional factor.

---

114. *Compare, e.g.*, *Wyeth*, 129 S. Ct. at 1192–94, *with id.* at 1211 (Thomas, J., concurring in judgment).
115. *See supra* section IV.B.
116. 529 U.S. 861 (2000).
117. *Id.* at 874–76.
118. *See, e.g.*, Altria Grp., Inc. v. Good, 129 S. Ct. 538, 551 (2008) (Thomas, J., dissenting, joined by Chief Justice Roberts and Justices Scalia and Alito) (arguing that the majority should have read the Federal Cigarette Labeling and Advertising Act's express preemption clause to preclude the plaintiff's state misrepresentative advertising claim).
119. See *Geier*, 529 U.S. at 897–98 (Stevens, J., dissenting, joined by Justices Souter, Thomas, and Ginsburg).
120. Hines v. Davidowitz, 312 U.S. 52, 67–68 (1941).

## C.  The Newcomers

One glaring limitation of this analysis is, of course, that it fails to account for the Court's recent additions: Chief Justice Roberts and Justices Alito, Sotomayor, and Kagan.  Such a defect is inevitable in any far-reaching analysis of a fluid adjudicative body,[121] but it is nonetheless important to recognize the limitation and make an effort to take into account the available data.  What analysis is possible suggests that these newcomers will have little effect on the Court's obstacle preemption decisions.  Thus far, Chief Justice Roberts and Justice Alito appear to be mirror images of Justice Scalia, voting with him in all three obstacle preemption decisions in which they have taken part.[122]  Perhaps surprisingly, in two of the three cases, their vote with Justice Scalia placed them in the opposite camp of Justice Kennedy,[123] the Court's most reliable pro–obstacle preemption vote.[124] This suggests that, like Justice O'Connor and Chief Justice Rehnquist, who preceded them, Chief Justice Roberts and Justice Alito will tend to vote in favor of obstacle preemption in contested cases, leaving the Court's balance of power unchanged.

Predictive data for Justices Sotomayor and Kagan are sparse.  Justice Sotomayor's preemption decisions prior to her elevation to the Supreme Court suggest, however, that like Justice Souter, she will side against preemption in the majority of contested cases.[125]  Also like

---

There is not—and from the very nature of the problem there cannot be—any rigid formula or rule which can be used as a universal pattern to determine the meaning and purpose of every act of Congress.  This Court, in considering the validity of state laws in the light of treaties or federal laws touching the same subject, has made use of the following expressions: conflicting; contrary to; occupying the field; repugnance; difference; irreconcilability; inconsistency; violation; curtailment; and interference.  But none of these expressions provides an infallible constitutional test or an exclusive constitutional yardstick.

   *Id.*

121.  This study attempts to limit the problem as much as possible by focusing on the period of stability following Justice Breyer's nomination to the Court in 1994.

122.  Haywood v. Drown, 129 S. Ct. 2108, 2118 (2009) (Thomas, J., dissenting) (dissenting from the majority's finding of preemption); Wyeth v. Levine, 129 S. Ct. 1187, 1217 (2009) (Alito, J., dissenting) (dissenting from the majority's finding of no preemption); Chamber of Commerce v. Brown, 554 U.S. 60 (2008) (voting with the majority in favor of preemption).

123.  *Haywood*, 129 S. Ct. 2108; *Wyeth*, 129 S. Ct. 1187.

124.  *See supra* section IV.B.

125.  *See* Dabit v. Merrill Lynch, 395 F.3d 25 (2d Cir. 2005), *vacated*, 547 U.S. 71 (2006); Empire Healthchoice Assurance, Inc. v. McVeigh, 396 F.3d 136 (2d Cir.

Justice Souter, she will employ the full panoply of interpretive tools to determine Congress's preemptive intent.[126] Like Justices Ginsburg, Stevens, and Breyer, for instance, she will presume against preemption where a plaintiff would be left with inadequate alternative remedies. In *Rombom v. United Airlines*,[127] where airline passengers brought suit against United Airlines for mental and physical distress after the airline summoned police officers to remove them from a plane, then-Judge Sotomayor considered relevant to a finding against field preemption both the inadequacy of administrative remedies, which would not have provided monetary damages, and the general presumption against incursion into areas of traditional state sovereignty.[128]

Justice Kagan's appointment to the Court presents the greatest puzzle. She, unlike other recent appointees, has no prior judicial experience and thus no body of decisions from which to predict the positions she will take as a member of the Court.[129] What little data are available suggest that Justice Kagan will, like Justice Stevens, her predecessor, tend to side with the Court's liberal wing on most issues. As U.S. Solicitor General, Kagan filed amicus briefs on behalf of the federal government in two significant preemption cases: *Williamson v. Mazda Motor of America, Inc.*[130] and *Chamber of Commerce v. Candelaria*.[131] Although we should be hesitant to read too much into positions taken while acting in a representative capacity, Kagan's stance in both briefs[132] places her squarely within the liberal camp.

At issue in *Williamson* is whether a safety regulation promulgated under the National Traffic and Motor Vehicle Safety Act of 1966

---

preempts a state tort law claim against Mazda Motor despite the statute's savings clause, which expressly preserves common-law liability.[133]  Mazda argued that, as in *Geier v. American Honda Motor Co.*,[134] the plaintiff's tort remedy is preempted, despite the savings clause, under a theory of obstacle preemption.[135]  In her amicus brief on behalf of the United States, then-Solicitor General Kagan rejected this view, distinguishing *Geier* and arguing against a finding of preemption.[136]  In so doing she resisted a broad reading of *Geier* and aligned herself with Justices Stevens, Souter, Thomas, and Ginsburg, all of whom dissented in *Geier*.[137]

By contrast, in *Candelaria*, a case challenging an Arizona's immigration statute,[138] Kagan's amicus brief argues in favor of federal preemption, but it does so under the theory most palatable to liberals,[139] express preemption, and in a case where federal preemption supports a liberal result—relaxation of anti-immigration measures.  In both cases, therefore, then-Solicitor General Kagan's position appears consistent with the Court's anti-obstacle preemption bloc.  If her positions as an advocate substantially reflect her own individual views, her replacement of Justice Stevens will not significantly alter the Court's balance on preemption questions.

On the whole, the available data suggest that recent changes to the Court's composition will not dramatically affect its obstacle preemption decisions.  A somewhat fragile but nonetheless important anti-obstacle preemption voting bloc remains, with Justices Sotomayor and Kagan replacing Justices Souter and Stevens, and that group continues to be counterbalanced by four relatively reliable votes from the Court's conservative wing in favor of obstacle preemption.

## VII.   CONCLUSION

This Article's analysis of the Court's recent preemption decisions yields several important results.  First, the analysis shows that, while the Court's preemption decisions considered in the aggregate may fall generally along ideological lines, predictive models based solely on judicial ideology miss an important determinant of legal decisions: doctrinal disputes among the Justices. Thus, while Justice Kennedy is generally regarded as a moderate, he shifts in a markedly conservative direction on preemption questions. Moreover, moving even further in the opposite direction, Justice Thomas switches poles entirely, from a strongly conservative position in the Court's decisions generally to a strong, traditionally liberal, view in obstacle preemption cases. Second, the analysis shows that Justice Thomas's renunciation of obstacle preemption in *Wyeth v. Levine* is both sincere and significant. His voting record backs his stated disdain for obstacle preemption, creating, when paired with the Court's liberal wing, a clearly defined but somewhat weak voting bloc against obstacle preemption. Third, the analysis shows that while the members of the anti-obstacle preemption voting bloc frequently arrive at the same non-preemptive results, they get there via quite different routes. Justice Thomas focuses on express congressional authorization to preempt, while the rest of the bloc members consider a wide variety of factors significant. Justice Thomas's use of such a blunt decision-making instrument creates a lack of flexibility that pushes him to the extreme, voting for obstacle preemption less than any other Justice, while the other Justices are able to tailor their analyses more closely to particular circumstances.

---

133.  Williamson v. Mazda Motor of Am., Inc., 84 Cal. Rptr. 3d 545, 547, 551 (Cal. Ct. App. 2008).

134.  529 U.S. 861 (2000).

135.  *See Williamson*, 84 Cal. Rptr. 3d at 552–53.

136.  *See* Brief for the United States as Amicus Curiae Supporting Petitioner, Williamson v. Mazda Motor of Am., No. 08-1314 (Aug. 6, 2010), 2010 WL 4150188.

137.  *Geier*, 529 U.S. at 886 (Stevens, J., dissenting).

138.  *See* Chicanos Por La Causa, Inc. v. Napolitano, 544 F.3d 976, 976–77 (9th Cir. 2008).

139.  The Court's conservatives find state law preempted under a theory of obstacle preemption as frequently or more frequently than they do under a theory of express preemption, while the Court's liberals exhibit the opposite tendency, finding obstacle preemption even less frequently than they do express preemption. *See supra* Table 1 & Figure A.

Considered collectively, these insights should allow court watchers to predict outcomes more reliably and potential litigants to target the arguments most likely to persuade each Justice. The Justices' views shift dramatically depending on the particular case's theory of preemption, and even Justices who consistently vote together do so for markedly different reasons.